\documentstyle[osa,manuscript]{revtex}  
\begin{document}                
\title{Block bond-order potential as a convergent moments-based method}
\author{T. Ozaki}
\address{Japan Advanced Institute of Science and Technology,
        Tatsunokuchi, Ishikawa 923-1292, Japan}
\author{M. Aoki} 
\address{Department of Electrical Electronic Engineering,
        Gifu University, 1-1 Yanagido, Gifu 501-11, Japan}
\author{D. G. Pettifor}
\address{Department of Materials, University of Oxford,
        Parks Road, Oxford OX1 3PH, United Kingdom}
\maketitle
\vspace{-5mm}
\begin{abstract}                
     The theory of a novel bond-order potential, which is based on
     the block Lanczos algorithm, is presented within an orthogonal
     tight-binding representation. The block scheme handles
     automatically the very different character of $\sigma$ and
     $\pi$ bonds by introducing block elements, which produces
     rapid convergence of the energies and forces within
     insulators, semiconductors, metals, and molecules. 
     The method gives the first convergent results for vacancies
     in semiconductors using a moments-based method with a low
     number of moments. Our use of the Lanczos basis simplifies the
     calculations of the band energy and forces, which allows 
     the application of the method to the molecular dynamics
     simulations of large systems. As an illustration of this
     convergent $O(N)$ method we apply the block bond-order
     potential to the large scale simulation of the deformation
     of a carbon nanotube.
\end{abstract}
\section{INTRODUCTION}

 To understand meso- and macro-scale phenomena from the atomistic level
 is an important subject in computer-aided materials modelling. This
 challenging study is not only intended as a realistic search for useful
 materials, but also for finding novel cooperative phenomena
 involving many atoms within large systems. Computer simulations of
 materials have inevitably promoted the development of efficient
 algorithms for dealing with long time-scale phenomena.
 These methods have developed via two different approaches.
 The first, based on the continuum mechanics, is a hybrid approach
 that combines the continuum mechanics with atomistic simulations.
 \cite{Gilson,Wlodek}
 The second is more directly applying the molecular dynamics (MD)
 simulation to large systems by reducing the computational effort.
 The progress in these two approaches will enable us to bridge
 micro- and macro-scale phenomena. In this paper we address the
 latter approach.

 Atomistic simulations should be founded on a quantum mechanical model
 in order to simulate a wide range of materials within a single
 framework, since the electronic structure determines the energy and
 the forces on atoms.
 The local-density approximation (LDA) to density functional
 theory\cite{Kohn,Sankey,Ab-TB} and semiempirical methods such
 as the tight-binding (TB) approximation\cite{Slater,Sutton} reduce
 the complicated quantum many body interaction in condensed matter
 to a single electron problem. The resultant theory has been applied
 to a variety of problems in materials science.
 However, it exceeds the capacity of modern computers to treat large
 systems that include thousands of atoms, using widely known methods
 for solving the single electron problem such as the
 conjugate-gradient method, since the computational effort scales
 as the third power of the system size.

 Therefore, several efficient methods with linear scaling algorithms have
 been proposed during the last decade.
 \cite{Pettifor,Aoki,Ozaki,Horsfield,Horsfield2,Goedecker,Goedecker2,Yang,Galli,Mauri,Mauri2,Li,Daw}
 These $O(N)$ methods can be roughly divided into two categories:
 variational methods and moments-based methods.
 The former are the density matrix (DM) methods\cite{Li,Daw}
 and the localized orbital (LO) methods\cite{Galli,Mauri,Mauri2}
 which lead to linear scaling algorithms from the locality of the density
 matrix.
 The latter include the bond-order potential (BOP) method 
 \cite{Pettifor,Aoki,Ozaki,Horsfield,Horsfield2} and
 the Fermi operator expansion (FOE) method\cite{Goedecker,Goedecker2}
 which are intrinsically linear in the scaling of
 the computational effort because the enegy and forces are expanded
 in a finite moment expansion. Several applications of these
 variational methods have already been performed for large systems,
 which have shown the power of these $O(N)$ methods.\cite{Galli2,Canning}
 However, several problems remain in these $O(N)$ methods.

 Firstly, it is well known that the variational $O(N)$ methods
 produce large errors in the energy of metallic systems with these
 long range correlations in the density matrix.\cite{Comparison}
 In these cases there is no justification for cutting the matrix
 elements off at short distances in the density matrix.
 Unfortunately, if the cutoff distance is increased to decrease the
 error in the energy, then the calculation effort increases significantly.

 Secondly, it is well documented that within moments-based methods,
 the vacancy in diamond or silicon can not be described within a low
 number of moments (about 20).\cite{Comparison,Kress}
 A very large number of moments (about
 200) is needed to reproduce the correct vacancy formation
 energy.\cite{Voter} In the BOP method the forces become exact as the bond
 orders converge to the exact values. This implies that the forces are not
 consistent with the total energy if the recursion is terminated
 at a finite number of levels. In the other moments-based methods,
 such as the FOE method\cite{Goedecker,Goedecker2} and the global
 density of states method,\cite{Horsfield2,GDOS} the exact forces can be
 calculated. However, these methods are also unable to reproduce the
 vacancy formation energy within a low number of
 moments.\cite{Comparison}

 Any robust $O(N)$ method should satisfy the following criteria.
 Firstly, the method should give accurate energies for a wide range
 of materials (insulators, semiconductors, metals, and molecules)
 with minimum computational effort. Secondly, the Hellmann-Feynman
 forces should be consistent with the total energy at any useful
 level of approximation. Thirdly, the algorithm should be suitable
 for parallel computation.

 Our goal is to establish the bond-order potential (BOP) method as 
 an $O(N)$ method that satisfies these three criteria.
 In Secs.~I\hspace{-0.2mm}I and I\hspace{-0.2mm}I\hspace{-0.2mm}I,
 we present the theory of the block BOP\cite{Ozaki} within the
 orthogonal tight-binding (TB) representation.
 We stress that the introduction of block elements into the BOP
 formalism improves remarkably the accuracy of the energy and forces.
 In Sec.~I\hspace{-0.2mm}V we explain the reason why the block BOP
 gives accurate energies with a low number of moments by analyzing
 the vacancy formation energy of diamond carbon in terms
 of the bond-order. In the remainder of this paper the deformation of
 a single-wall carbon nanotube is used to demonstrate the
 applicability of the method to large scale atomistic simulations.

\section{THEORY}

\subsection{Tight-binding}

 We develop the block BOP within the two center orthogonal TB
 representation.\cite{Sutton,Heine}
 It will be assumed that the basis set is an orthonormal set of atomiclike
 orbitals $\vert i\alpha\rangle$ where $i$ is a site index, and $\alpha$
 an orbital index. The Hamiltonian can be represented by the matrix
 $H_{i\alpha,j\beta}=\langle i\alpha\vert \hat{H}\vert j\beta\rangle$.
 The on-site elements of the matrix are written as $\epsilon_{i\alpha}$.
 The cohesive energy, assuming that the electrons are at a finite temperature
 $T$, is the sum of bond, promotion, and repulsive energies:
 \begin{eqnarray}
     E_{\rm coh} & = & E_{\rm bond} + E_{\rm prom} + E_{\rm rep},
 \end{eqnarray}
 where the repulsive energy is given by the sum of pair potentials or
 embedded potentials which are usually determined so that the TB
 model reproduces equilibrium structures and elastic constants.
 The bond energy is the attractive contribution that leads to cohesion.
 There are two different but equivalent expressions that describe
 the bond energy. The first gives the bond energy in terms of
 the {\it on-site} density of states as follows:
 \begin{eqnarray}
     E_{\rm bond} & = & 2\sum_{i\alpha}\int (E-\epsilon_{i\alpha})
                  n_{i\alpha}(E)f\left(\frac{E-\mu}{k_BT}\right)dE,
 \end{eqnarray}
 where $n_{i\alpha}(E)$ is the density of states projected onto orbital
 $\vert i\alpha\rangle$, and the function $f(x)=1/[1+{\rm exp}(x)]$ is
 the Fermi function. The second gives the bond energy explicitly
 in terms of the individual {\it intersite} bond energies as
 follows:
 \begin{eqnarray}
     E_{\rm bond} & = & \frac{1}{2}
                    \sum_{i\alpha\ne j\beta}
                    \Bigl(2\Theta_{i\alpha,j\beta}H_{j\beta,i\alpha}
                    \Bigr),
 \end{eqnarray}
 where $\Theta_{i\alpha,j\beta}$ is the bond-order between orbitals
 $\vert i\alpha\rangle$ and $\vert j\beta\rangle$, and the expression
 parenthesis represents the corresponding bond energy associated
 with orbitals $\vert i\alpha\rangle$ and $\vert j\beta\rangle$.
 This allows us to interpret the bonding and structure of molecules
 and solids from a chemical point of view.\cite{Pettifor-Book}
 It should be noted that the bond-order is not pairwise but 
 is determined by the particular arrangement and connectivity
 of the atoms adjacent to the two atoms forming the bond.
 In the block BOP representation the two different expressions
 Eqs.~(2) and (3) for the bond energy are exactly identical at any
 level of approximation. The proof will be given later subsection.
 The promotion energy is defined by
 \begin{eqnarray}
     E_{\rm prom} & = & \sum_{i\alpha}
                    \left(\epsilon_{i\alpha}N_{i\alpha}
                    -\epsilon^0_{i\alpha}N_{i\alpha}^0\right),
 \end{eqnarray}
 where $N_{i\alpha}$ and $N_{i\alpha}^0$ are the number of electrons in 
 $\vert i\alpha\rangle$ in the condensed and free atomic systems, respectively.
 The promotion energy is repulsive due to the excitation of electrons
 from their free atomic ground state as the atoms are brought together.
 Therefore, the cohesive energy of a system is determined by the balance
 between the attractive bond energy and the repulsive pairwise/embedding
 and promotion energies. The bond and promotion energies can be repartitioned
 into the band and atomic energies:
 \begin{eqnarray}
     \nonumber
     E_{\rm bond} + E_{\rm prom} & = &
                    \sum_{i\alpha\ne j\beta}
                    \Theta_{i\alpha,j\beta} H_{j\beta,i\alpha}
                     + 
                    \sum_{i\alpha}
                    \left(\epsilon_{i\alpha}N_{i\alpha}
                    -\epsilon^0_{i\alpha}N_{i\alpha}^0\right)\\
      \nonumber
                         & = &
                    \sum_{i\alpha,j\beta}
                    \Theta_{i\alpha,j\beta} H_{j\beta,i\alpha}
                     - 
                    \sum_{i\alpha}
                     \epsilon^0_{i\alpha}N_{i\alpha}^0\\
                         & = &
                          E_{\rm band} - E_{\rm atoms}.
 \end{eqnarray}
 $E_{\rm band}$ is equal to the energy which is defined by integrating
 $\sum_{i\alpha}E n_{i\alpha}(E)$ up to the Fermi level.

 In the TB model the single particle eigenfunctions are expanded in
 a basis set that is an orthonormal set of real atomiclike orbitals:
 $\vert i\alpha\rangle$.
 \begin{eqnarray}
     |\phi\rangle = \sum_{i\alpha}C_{i\alpha}^{(\phi)}|i\alpha\rangle,
 \end{eqnarray}
 where the expansion coefficients are defined by
 $C_{i\alpha}^{(\phi)}\equiv \langle i\alpha\vert\phi\rangle$. 
 $C_{i\alpha}^{(\phi)}$ is always real because of real atomic orbitals
 and Hamiltonian. Then the bond-orders may be defined in terms of 
 the expansion coefficients as follows:
 \begin{eqnarray}
     \Theta_{i\alpha,j\beta} =
       2\sum_{\phi}C_{j\beta}^{(\phi)}C_{i\alpha}^{(\phi)}
       f\left(\frac{\epsilon^{(\phi)}-\mu}{k_BT}\right),
 \end{eqnarray}
 where the factor 2 accounts for spin degeneracy. $\epsilon^{(\phi)}$ is 
 the eigenvalue corresponding to an eigenstate $\vert\phi\rangle$.
 
 The force on atom $k$ is obtained by differentiating Eq.~(1) with respect to
 atomic positions:
 \begin{eqnarray}
     \nonumber
     {\bf F}_k & = & -\frac{\partial E_{\rm coh}}{\partial {\bf r}_k}\\
               & = &
       -\sum_{i\alpha,j\beta}
       \left(
        \frac{\partial \Theta_{i\alpha,j\beta}}{\partial {\bf r}_k}
         H_{j\beta,i\alpha}
         +
        \Theta_{i\alpha,j\beta}
        \frac{\partial H_{j\beta,i\alpha}}{\partial {\bf r}_k}
       \right)
      - \frac{\partial E_{\rm rep}}{\partial {\bf r}_k}.
 \end{eqnarray}
 The first term of Eq.~(8) is identically zero so that 
 \begin{eqnarray}
     {\bf F}_k  = 
      -\sum_{i\alpha,j\beta}
        \Theta_{i\alpha,j\beta}
        \frac{\partial H_{j\beta,i\alpha}}{\partial {\bf r}_k}
      - \frac{\partial E_{\rm rep}}{\partial {\bf r}_k},
 \end{eqnarray}
 where the first term of Eq.~(9) is the Hellmann-Feynman force.
 If the bond-orders are approximate values, then the sum of the
 derivatives of the bond-orders with respect to atomic positions
 will not be zero, so that Eq.~(8) gives the exact force which is
 consistent with the total energy. However, in the block BOP
 representation the forces are given by Eq.~(9), since it is very
 difficult to evaluate the derivatives of the bond-orders. Hence,
 the forces calculated by block BOP become exact as the bond-orders
 converge to the exact values. In Sec.~2 the compatibility between
 the force and the energy will be discussed from numerical tests
 using constant energy molecular dynamics simulations.

 Although the forces are not consistent with the total energy in 
 the usual BOP methods, it is possible to evaluate the exact forces
 at any level of approximate by the other moments-based method,
 global density of states method. \cite{GDOS}
 However the use of the global moments, which are introduced to
 decrease the computational effort, leads to a reduced rate of
 convergence of the energy as a function of the number of
 moments. In Appendix of this paper we present a novel
 method to evaluate the exact forces.

\subsection{Block bond-order potential}

 The local density of states and bond-orders can be related to the one
 particle Green's functions. The one particle Green's function
 operator is defined by
 \begin{eqnarray}
    \nonumber
    \hat{G}(Z) & = & (Z-\hat{H})^{-1}\\
               & = & \sum_{\phi}
                   \frac{|\phi\rangle\langle \phi|}{Z-\epsilon^{(\phi)}}.
 \end{eqnarray}
 Then the imaginary part of the diagonal elements of the Green's function
 matrix give the local density of states:
 \begin{eqnarray}
    \nonumber
    {\rm Im}~G_{i\alpha,i\alpha}(E+{\rm i0^+})
               & = &
                  \sum_{\phi}
               \frac{-0^+\langle i\alpha|\phi\rangle\langle \phi|i\alpha \rangle}
                    {(E-\epsilon^{(\phi)})^2+(0^+)^2}\\
    \nonumber
               & = & 
                  -\pi \sum_{\phi}
                    (C_{i\alpha}^{(\phi)})^2\delta(E-\epsilon^{(\phi)})\\
    \nonumber
               & = & 
                -\pi n_{i\alpha}(E).
 \end{eqnarray}
 Therefore
 \begin{eqnarray}
   n_{i\alpha}(E)
                =  
           -\frac{1}{\pi}{\rm Im}~G_{i\alpha,i\alpha}(E+{\rm i0^+}),
 \end{eqnarray}
 where $G_{i\alpha,i\alpha}(Z)=
 \langle i\alpha\vert \hat{G}(Z)\vert i\alpha\rangle$, $0^+$
 represents a positive infinitesimal, and $\delta(x)$ is the delta
 function. The imaginary part of the off-diagonal elements of the
 Green's function matrix has the following relation to the
 expansion coefficients of the single particle eigenfunctions:
 \begin{eqnarray}
    {\rm Im}~G_{i\alpha,j\beta}(E+{\rm i0^+})
                =  
                  -\pi \sum_{\phi}
                    C_{j\beta}^{(\phi)}C_{i\alpha}^{(\phi)}
                    \delta(E-\epsilon^{(\phi)}).
 \end{eqnarray}
 Multiplying the both sides of Eq.~(12) by the Fermi function,
 integrating with respect to the energy we obtain the following
 useful expression for the bond-order:
 \begin{eqnarray}
    \nonumber
    \lefteqn{{\rm Im}\int G_{i\alpha,j\beta}(E+{\rm i0^+})
                f(\frac{E-\mu}{k_BT})dE}\\
       \nonumber   && = 
                  -\pi \sum_{\phi}
                    C_{j\beta}^{(\phi)}C_{i\alpha}^{(\phi)}
                    \int\delta(E-\epsilon^{(\phi)})
                    f(\frac{E-\mu}{k_BT})dE\\
       \nonumber   && = 
                  -\pi \sum_{\phi}
                    C_{j\beta}^{(\phi)}C_{i\alpha}^{(\phi)}
                    f(\frac{\epsilon^{(\phi)}-\mu}{k_BT})\\
       \nonumber   && =
                 -\frac{\pi}{2}\Theta_{i\alpha,j\beta}.
 \end{eqnarray}
 Therefore
 \begin{eqnarray}
    \Theta_{i\alpha,j\beta} = -\frac{2}{\pi}
                {\rm Im}\int G_{i\alpha,j\beta}(E+{\rm i0^+})
                f(\frac{E-\mu}{k_BT})dE.
 \end{eqnarray}
 The evaluations of the bond energy Eqs.~(2) and (3) require calculating the
 local density of states and bond-orders. We obtain the local density of
 states and bond-orders from the Green's function through Eqs.~(11) and (13).
 The diagonal elements of the Green's function matrix can be calculated in a
 numerically stable way by the recursion method.\cite{Haydock,Haydock2}
 Block BOP is a general recursion method for evaluating
 efficiently both the diagonal and off-diagonal elements of the
 Green's function matrix by the recursion method.
 The first step of the recursion method is to tridiagonalize the Hamiltonian
 using the Lanczos algorithm.\cite{Lanczos} In the block BOP we
 introduce the {\it block} Lanczos algorithm with the starting state as a
 single site containing all the valence orbitals rather than the
 usual {\it scalar} Lanczos algorithm with a single starting
 orbital.\cite{Ozaki} However, the application of the conventional block 
 algorithm\cite{Jones,Inoue} to finite systems such as molecules 
 introduces a numerical instability, since the terminal number of
 recursion levels of the $\pi$ bond are different from that of
 the $\sigma$ bond in the recursive algorithm.
 Therefore, we modify the conventional block Lanczos algorithm.
 A series of procedures for the modified block Lanczos
 algorithm can be carried out as follows:
 \begin{eqnarray}
    \vert U_0) = (\vert i1\rangle,\vert i2\rangle,\dots,\vert iM_i\rangle ).
 \end{eqnarray}
 \begin{eqnarray}
    \underline{A}_n  =  (U_n\vert\hat{H}\vert U_n).
 \end{eqnarray}
 \begin{eqnarray}
    \vert r_n)       = 
            \hat{H}\vert U_{n})-\vert U_{n-1})\hspace{0.4mm}
            ^t\hspace{-0.4mm}\underline{B}_{n}
            -\vert U_{n})\underline{A}_{n}.
 \end{eqnarray}
 \begin{eqnarray}
    (\underline{B}_{n+1})^2 = (r_n\vert r_n).
 \end{eqnarray}
 \begin{eqnarray}
    (\underline{\lambda}_{n})^2 =
                  \hspace{0.4mm}^t\hspace{-0.3mm}\underline{V}_n
                                (\underline{B}_{n+1})^2
                                \underline{V}_n.
 \end{eqnarray}
 \begin{eqnarray}
    \underline{B}_{n+1} = \underline{\lambda}_{n}\hspace{0.3mm}
                      \hspace{-0.4mm}^t\underline{V}_n.
 \end{eqnarray}
 \begin{eqnarray}
    (\underline{B}_{n+1})^{-1} =
      \underline{V}_n\underline{\lambda}_{n}^{-1}.
 \end{eqnarray}
 \begin{eqnarray}
    \vert U_{n+1}) = \vert r_n)(\underline{B}_{n+1})^{-1}.
 \end{eqnarray}
 $\underline{A}_n$ and $\underline{B}_n$ are recursion block coefficients with
 $M_i\times M_i$ in size, where $M_i$ is the number of atomic orbitals on the
 starting atom $i$, and the underline indicates that the element is a block.

 The states $\vert U_n)=(\vert L_{n1}\rangle,\vert L_{n2}\rangle,
 \cdots,\vert L_{nM_i}\rangle)$ represent the Lanczos basis, and are
 orthonormal and block-tridiagonalize the Hamiltonian.
 The modified algorithm gives different expressions for the
 block elements $\underline{B}_{n+1}$ and these inverses compared with
 the conventional algorithm. The block elements in the conventional block
 Lanczos algorithm are defined by
 \begin{eqnarray}
    \underline{B}_{n+1}
         = \underline{V}_n\underline{\lambda}_{n}\hspace{0.4mm}
             ^t\hspace{-0.4mm}\underline{V}_n.
 \end{eqnarray}
 \begin{eqnarray}
    (\underline{B}_{n+1})^{-1}
    = \underline{V}_n\underline{\lambda}_{n}^{-1}\hspace{0.4mm}
               {^t\hspace{-0.4mm}\underline{V}_n}.
 \end{eqnarray}
 The failure in the conventional algorithm can be illustrated by a
 carbon trimer with a linear chain structure along the $x$-axis.
 If the block Lanczos algorithm is applied with the central atom in
 the trimer as the starting state, then the $p_y$ and
 $p_z$ orbitals span two independent subspaces.
 Thus, the recursive algorithm finishes after only one iteration for the
 Lanczos vectors concerned with the $p_y$ and $p_z$ orbitals.
 This gives two zero eigenvalues in the four eigenvalues of the block
 element $(\underline{B}_2)^2$. Then one can not evaluate the inverse
 of $\underline{B}_2$ using Eq.~(23).
 Therefore, defining $\underline{B}_2$ and its inverse by the modified
 Eqs.~(20) and (21), respectively, and assuming that the diagonal
 elements of $\underline{\lambda}_1^{-1}$ corresponding to the zero
 eigenvalues are zero we have
 \begin{eqnarray}
   \underline{B}_2(\underline{B}_2)^{-1}=
             \left(
               \begin{array}{cccc}
                 1 &   &   &   \\
                   & 1 &   &   \\
                   &   & 0 &   \\
                   &   &   & 0 \\
               \end{array}
             \right).
 \end{eqnarray}
 \begin{eqnarray}
  (U_2\vert U_2) =
             \left(
               \begin{array}{cccc}
                 1 &   &   &   \\
                   & 1 &   &   \\
                   &   & 0 &   \\
                   &   &   & 0 \\
               \end{array}
             \right).
 \end{eqnarray}
 $\vert U_2)$ is reduced to the state with two vectors, while the starting
 state $\vert U_0)$ is constructed by the four vectors, which permits us to
 iterate once more with the recursive algorithm. The conventional block Lanczos
 algorithm does not satisfy both Eqs.~(24) and (25), since the block elements
 $\underline{B}_2$ and the inverse are obtained from the unitary
 transformations of $\underline{\lambda}_1$ and the inverse, respectively.
 Therefore, the conventional algorithm terminates at this recursion level
 even though the Lanczos vectors for the $\sigma$ orbital can still hop.
 This reduction of the state avoids the numerically instabilities
 for the case of small eigenvalues of $(\underline{B}_{n+1})^2$, 
 even when the eigenvalues are not zero. 

 Application of the block Lanczos algorithm defines an orthonormal basis set 
 called the Lanczos vector or basis. The Lanczos vectors reflect the
 neighboring atomic arrangement of the starting site. In Fig.1 we show
 the Lanczos vectors on an $s$-valent square lattice. The Lanczos vectors
 spread gradually from the central atom as the number of recursion
 levels increases. Thus, we now expand a one electron
 eigenstate using the Lanczos vectors:
 \begin{eqnarray}
    \vert \phi\rangle = \sum_{n\nu}D_{n\nu}^{(\phi)}\vert L_{n\nu}\rangle,
 \end{eqnarray}
 where $D_{n\nu}^{\phi}\equiv\langle L_{n\nu}\vert\phi\rangle$.
 Then the representation based on the atomic basis can be transformed into that
 of the Lanczos basis set by the matrix $U$ such that
 \begin{eqnarray}
    T^L =\hspace{0.4mm} ^t\hspace{-0.4mm}U T U,
 \end{eqnarray}
 where $U$ is defined by $\langle i\alpha\vert L_{n\nu}\rangle$, and T can be the
 Hamiltonian $H$, the derivative of Hamiltonian with respect to atomic position
 $\partial H/ \partial {\bf r}_i$, the bond-order $\Theta$, or the Green's
 function $G(Z)$ matrix. The index $L$ indicates the representation based on
 the Lanczos basis. Equation~(27) is a pseude unitary transformation, and
 the matrix $U$ becomes unitary when the number of the recursion
 levels is infinity in infinite systems.
 If the block Lanczos algorithm is started through Eq.~(14) with
 the atomic orbitals on atom $i$ as the starting state, then considering
 Eq.~(27) and the orthonormality of the Lanczos basis, we can relate the bond
 orders in the Lanczos basis representation to the bond-orders based on the
 atomic basis by the following simple relation:
 \begin{eqnarray}
    \underline{\Theta}_{ij} =
          \sum_n \underline{\Theta}^L_{0n}
                 {^t\hspace{-0.4mm}\underline{U}_{nj}}, 
 \end{eqnarray}
 where $\underline{\Theta}_{ij}$ and $\underline{\Theta}_{0n}^L$ are 
 the block elements of the bond-orders for the atoms $i$ and $j$, and the
 states $\vert U_{0})$ and $\vert U_{n})$, respectively.
 For example $\underline{\Theta}_{ij}$ signifies
 \begin{eqnarray}
    \underline{\Theta}_{ij} =
     \left(
       \begin{array}{cccc}
         \Theta_{i1,j1} & \Theta_{i1,j2} & \cdots & \Theta_{i1,jM_j}\\
         \Theta_{i2,j1} & \Theta_{i2,j2} & \cdots & \Theta_{i2,jM_j}\\
         \multicolumn{4}{c}{\dotfill}\\
         \Theta_{iM_i,j1} & \Theta_{iM_i,j2} & \cdots & \Theta_{iM_i,jM_j}\\
       \end{array}
     \right),
 \end{eqnarray}
 where $M_i$ and $M_j$ are the numbers of atomic orbitals including atoms $i$ and
 $j$, respectively.
 In Eq.~(28) $^t\hspace{-0.4mm}\underline{U}_{nj}$, which is the (n,j)
 block element of the matrix $^t\hspace{-0.4mm}U$, is defined by
 \begin{eqnarray}
    ^t\hspace{-0.4mm}\underline{U}_{nj} =
     \left(
       \begin{array}{cccc}
         \langle L_{n1}\vert j1\rangle &
         \langle L_{n1}\vert j2\rangle &
          \cdots & \langle L_{n1}\vert jM_j\rangle\\
         \langle L_{n2}\vert j1\rangle &
         \langle L_{n2}\vert j2\rangle &
         \cdots & \langle L_{n2}\vert jM_j\rangle \\
         \multicolumn{4}{c}{\dotfill}\\
         \langle L_{nM_i}\vert j1\rangle &
         \langle L_{nM_i}\vert j2\rangle &
         \cdots & \langle L_{nM_i}\vert jM_j\rangle \\
       \end{array}
     \right).
 \end{eqnarray}
 The simple relation Eq.~(28) allows us to evaluate the bond-order in terms
 of the Lanczos basis representation. We have only to calculate the 0th block
 line, which are the bond-orders between the starting atom and the Lanczos
 vectors surrounding the atom, of the bond-order matrix.
 In the block BOP the bond-orders are evaluated in the Lanczos basis
 representation, and then we get the bond-orders based on the atomic basis
 from Eq.~(28).

 It is essential to start the block Lanczos algorithm with a single site
 as in Eq.~(14). Although it is possible to derive an analogous
 transformation to Eq.~(28) using the usual scalar Lanczos
 algorithm, the bond energy of the system depends on
 the rotation of the system.\cite{Inoue} Thus, the use of the scalar algorithm
 is not appropriate, since the bond energy should be invariant to the rotation
 of the system. We could also start the recursion with a cluster
 containing a neighbor shell of atoms instead of a single site.\cite{Jones}
 However, this choice is unsuitable because it is highly
 computationally intensive.

 In the Lanczos representation the Hamiltonian is block-tridiagonalized:
 \begin{eqnarray}
    (U_m\vert\hat{H}\vert U_n) = \left\{
            \begin{array}{ll}
               \underline{A}_n & \quad  \mbox{if $m=n$},\\
    \hspace{-1.0mm}   ^t\hspace{-0.4mm}\underline{B}_n  & \quad \mbox{if $m=n-1$},\\
               \underline{B}_{n+1} & \quad  \mbox{if $m=n+1$},\\
               \underline{0}      & \quad  \mbox{otherwise}.
            \end{array}\right.
 \end{eqnarray}
 The block element $\underline{G}_{00}(Z)=(U_{0}\vert \hat{G}\vert U_{0})$
 can be written explicitly by the form of the multiple inverse, since the
 Green's function matrix $G(Z)$ is the inverse of the matrix $(Z{\rm I}-H)$.
 Appling repeatedly the partitioning method, which is a method for calculating
 the inverse of matrices, to the matrix $(Z{\rm I}-H)$ we get
 \begin{eqnarray}
   \underline{G}^L_{00}(Z)
        =[Z\underline{\rm I}-\underline{A}_0-\hspace{0.4mm}^t\hspace{-0.4mm}\underline{B}_1[
          Z\underline{\rm I}-\underline{A}_1-\hspace{0.4mm}^t\hspace{-0.4mm}\underline{B}_2[
                       \cdots
          ]^{-1}\underline{B}_2
          ]^{-1}\underline{B}_1
          ]^{-1}.
 \end{eqnarray}
 $\underline{G}_{00}^{L}(Z)$ is equal to the block element
 $\underline{G}_{ii}(Z)$ based on the atomic basis, since we have started the
 block Lanczos algorithm with Eq.~(14). Therefore, the local density of states
 can be evaluated from the diagonal elements by Eq.~(11). Also the trace of 
 $\underline{G}_{00}^{L}(Z)$ gives the local density of states on atom $i$.

 Moreover, by taking account of the block-tridiagonalized Hamiltonian
 and the identity $(Z{\rm I}-H)G(Z)={\rm I}$~ in the Lanczos basis representation,
 the off-diagonal elements of the Green's function matrix
 $\underline{G}_{0n}^L$ may be obtained from the following recurrence
 relation:
 \begin{eqnarray}
      \underline{G}^{L}_{0n}(Z)
      =
    \biggl(
      \underline{G}^{L}_{0n-1}(Z)(Z\underline{\rm I}
       -\underline{A}_{n-1})
       -\underline{G}^{L}_{0n-2}(Z)
         \hspace{0.4mm}^t\hspace{-0.4mm}\underline{B}_{n-1}
         -\delta_{1n}\underline{\rm I}
    \biggr)
         (\underline{B}_{n})^{-1},
 \end{eqnarray}
 where $\delta$ is the Kronecker's delta, $\underline{G}_{0-1}(Z)$ and 
 $\hspace{0.1mm}^t\hspace{-0.4mm}\underline{B}_{0}$
 are $\underline{0}$, respectively. All the off-diagonal
 block elements $\underline{G}_{0n}^L(Z)$ are related to the diagonal block
 element $\underline{G}_{00}^L(Z)$. Once $\underline{G}_{00}^L(Z)$ has been
 obtained, the off-diagonal block elements are easily evaluated from the
 above recursive relation. The simplicity of evaluating the off-diagonal
 block elements is an important advantage of the Lanczos basis
 representation. The block elements of the Green's function matrix have the
 same relation to the bond-orders based on the Lanczos basis as that
 of the atomic basis representation:
 \begin{eqnarray}
    \underline{\Theta}_{0n}^L
                  =
                -\frac{2}{\pi}
                {\rm Im}\int 
                \underline{G}_{0n}(E+{\rm i0^+})
                f(\frac{E-\mu}{k_BT})dE
 \end{eqnarray}

 In case the bond-orders are evaluated by Eqs.~(28) and (34), we can prove that
 the two different expressions Eqs.~(2) and (3) for the bond energy are
 identical at any level of approximation.
 Consider the trace of $G(Z)(Z{\rm I}-H)$. Transforming the trace of the
 atomic basis representation into that of the Lanczos basis using Eq.~(27),
 and making use of the identity $G(Z)(Z{\rm I}-H)={\rm I}$ in the Lanczos
 basis representation we see that the trace is a constant:
 \begin{eqnarray}
   \nonumber
   \lefteqn{
     {\rm tr}\left\{G(Z)(Z{\rm I}-H)\right\}
   }\\
  \nonumber
   && =
      \sum_{i}{\rm tr}\left\{Z\underline{G}_{ii}(Z)\right\}      
     -\sum_{ij}{\rm tr}\left\{\underline{G}_{ij}(Z)\underline{H}_{ji}\right\}\\
  \nonumber
   && = 
      \sum_{i}{\rm tr}\left\{Z\underline{G}^{L^{(i)}}_{00}(Z)\right\}
     -\sum_{in}{\rm tr}\left\{
                \underline{G}^{L^{(i)}}_{0n}(Z)
                \underline{H}^{L^{(i)}}_{n0}
               \right\}\\
   && = 
      \sum_{i} 
      {\rm tr}({\rm \underline{I}}^{(i)}),
 \end{eqnarray}
 where $\underline{{\rm I}}_{i}$ is a unit matrix with $M_i\times M_i$ in size.
 The index $L^{(i)}$ indicates the representation based on the Lanczos
 basis with the starting state on atom $i$. Considering the imaginary parts
 of the trace we have
 \begin{eqnarray}
        {\rm Im}
         \sum_{i\alpha} 
         ZG_{i\alpha,i\alpha}(Z) 
    =
     {\rm Im}
     \sum_{i\alpha,j\beta}
      G_{i\alpha,j\beta}(Z)H_{j\beta,i\alpha}.
 \end{eqnarray}
 We see that the two expression for the bond energy give the same energy,
 since the Green's functions can be related to the local density of states
 and bond-orders through Eqs.~(11) and (13), respectively.
 The block BOP, thus, provides the equivalence of the two expressions for
 the bond energy in a natural way, whereas in the usual BOP the
 Green's functions need a carefully chosen truncator in order to satisfy
 the sum rule.\cite{Aoki}

\subsection{Moment description}

 The moments of the local density of states allow us to link the behavior
 of the electronic structure to the local topology about the given site.
 \cite{Horsfield,Horsfield2,Pettifor-Book}  We now discuss the
 relation between the block recursion matrices and the moments of 
 the density of states.
 From Eq.~(10) for $\vert Z\vert \to \infty$, the diagonal element
 $\underline{G}_{00}^{L}(Z)$ can be rewritten as follows:
 \begin{eqnarray}
    \nonumber
    \underline{G}_{00}^L(Z)
    & = &
       \sum_{\phi}
              \frac{(U_0\vert\phi\rangle\langle \phi\vert U_0)}
                  {Z-\epsilon^{(\phi)}}\\
    \nonumber
    & = & 
       \sum_{\phi}\underline{d}_{00}^{(\phi)}
       \left(
          \sum_{p=0}^{\infty}
           \frac{(\epsilon^{(\phi)})^p}{Z^{p+1}}
       \right)\\
    & = & 
       \sum_{p=0}^{\infty}
        \frac{\underline{\mu}_{00}^{(p)}}
          {Z^{p+1}},    
 \end{eqnarray}
 where
 \begin{eqnarray}
    \underline{d}_{00}^{(\phi)} =
     \left(
       \begin{array}{cccc}
         D_{i1}^{(\phi)}D_{i1}^{(\phi)}
       & D_{i2}^{(\phi)}D_{i1}^{(\phi)}
       & \cdots
       & D_{ip}^{(\phi)}D_{i1}^{(\phi)}\\
         D_{i1}^{(\phi)}D_{i2}^{(\phi)}
       & D_{i2}^{(\phi)}D_{i2}^{(\phi)}
       & \cdots
       & D_{ip}^{(\phi)}D_{i2}^{(\phi)}\\
         \multicolumn{4}{c}{\dotfill}\\
         D_{i1}^{(\phi)}D_{ip}^{(\phi)}
       & D_{i2}^{(\phi)}D_{ip}^{(\phi)}
       & \cdots
       & D_{ip}^{(\phi)}D_{ip}^{(\phi)}\\
       \end{array}
     \right),
 \end{eqnarray}
 \begin{eqnarray}
    \underline{\mu}_{00}^{(p)}
    & = &
    \sum_{\phi}
    \underline{d}_{00}^{(\phi)}
    \left(\epsilon^{(\phi)}\right)^p,
 \end{eqnarray}
 and $\underline{\mu}_{00}^{(p)}$ is the block element of the $p$th moment
 for the atom $i$, the diagonal elements of which give the $p$th moments of
 the projected density of states $n_{i\alpha}(E)$. Thus, Eq.~(37) is the moment
 expansion of the Green's function $\underline{G}_{00}^{L}(Z)$.
 Also the $p$th block moment can be evaluated explicitly as the expectation
 value of the $p$th power of the Hamiltonian in terms of the block elements
 $\underline{A}_{n}$, $\underline{B}_{n}$:
 \begin{eqnarray}
    \nonumber
    \underline{\mu}_{00}^{(p)}
    & = &
     (U_0\vert \hat{H}^p \vert U_0)\\
    & = & 
     \sum_{m_1\cdots m_{p-1}}  
       (U_0\vert \hat{H}\vert U_{m_1})
       (U_{m_1}\vert \hat{H}\vert U_{m_2})
        \cdots 
       (U_{m_{p-1}}\vert \hat{H}\vert U_{0}).
 \end{eqnarray}
 The first few block moments are
 \begin{eqnarray}
    \nonumber
    \underline{\mu}_{00}^{(0)}& = & \underline{\rm I},\\
    \nonumber
    \underline{\mu}_{00}^{(1)}& = &\underline{A}_0,\\
    \underline{\mu}_{00}^{(2)}& = & (\underline{A}_0)^2
              +\hspace{0.4mm} ^t\hspace{-0.4mm}\underline{B}_1
                        \underline{B}_1.
 \end{eqnarray}
 From Eq.~(40) we see that the $p$th moment is the sum over all
 self-returning paths of length $p$.
 The first moment corresponds to a hop on a single site, the second to nearest
 neighbors and back, and so on. Thus, the atomic connectivity can be related 
 directly to the electronic structure through the description of the Green's
 function by the moments.

 Multiplying both sides of Eq.~(37) by $(E+0^+)^r$, and integrating 
 with respect to the energy $E$ we get the following relation:
 \begin{eqnarray}
    -\frac{1}{\pi}{\rm Im}  
     \int_{-\infty}^{\infty}E^r
     \underline{G}_{00}^L(E+0^+)dE
      =\underline{\mu}_{00}^{(r)}. 
 \end{eqnarray}
 This relation means that the imaginary part of the moment of the block
 diagonal element in the Green's function matrix is equal to the moment of
 the Hamiltonian.

 Let us define the orthogonal block polynomials $\underline{P}_n(x)$:
 \begin{eqnarray}
     x\underline{P}_n(x)
             =  \underline{P}_n(x)\underline{A}_n
 + \underline{P}_{n-1}(x)\hspace{0.4mm}^t\hspace{-0.4mm}\underline{B}_n 
 + \underline{P}_{n+1}(x)\underline{B}_{n+1},
 \end{eqnarray} 
 where $\underline{P}_{-1}(x)$ and $\underline{P}_{0}(x)$ are the zero matrix
 $\underline{0}$ and the the unit matrix $\underline{\rm I}$ with $M_i \times M_i$
 in size. By using the block polynomials the recursion block elements
 $\underline{A}_n$ and $\underline{B}_n$ can be expanded with the moments:
 \begin{eqnarray}
    \nonumber
    \underline{A}_{n} & = & (U_n\vert H \vert U_n)\\
    \nonumber
                      & = & 
         \hspace{0.4mm}^t\hspace{-0.4mm}\underline{P}_n(\hat H)(U_0\vert \hat{H}
                  \vert U_0)\underline{P}_n(\hat H)\\
                      & = &
                     \sum_m^{2n+1}
                        \underline{a}_m  
                        \underline{\mu}_{00}^{(m)}
                        \underline{a'}_m.
 \end{eqnarray}
 \begin{eqnarray}
    \nonumber
    \underline{B}_{n} & = & (U_n\vert H \vert U_{n-1})\\
    \nonumber
                      & = & 
         \hspace{0.4mm}^t\hspace{-0.4mm}\underline{P}_n(\hat H)(U_0\vert \hat{H}
                           \vert U_0)\underline{P}_{n-1}(\hat H)\\
                      & = &
                     \sum_m^{2n} 
                        \underline{b}_m
                        \underline{\mu}_{00}^{(m)}
                        \underline{b'}_m.
 \end{eqnarray} 
 In the derivations of Eqs.~(44) and (45) we have assumed the substitution:
 $\vert U_0)\hat{H} \to \hat{H}\vert U_0)$ and
 $\hat{H}(U_0\vert \to (U_0\vert\hat{H}$.
 The block coefficients $\underline{a}_m$, $\underline{a}'_m$,
 $\underline{b}_{m}$, and $\underline{b}'_{m}$ are given by the recursion 
 block elements. For example $\underline{A}_1$ and $\underline{B}_1$ can be
 written as follows:
 \begin{eqnarray}
     \underline{A}_{1} = 
     (\hspace{0.4mm}^t\hspace{-0.4mm}\underline{B}_1)^{-1}
     \left\{
          \underline{\mu}_{00}^{(3)}
        - \underline{A}_0\underline{\mu}_{00}^{(2)}
        - \underline{\mu}_{00}^{(2)}\underline{A}_0
        + \underline{A}_0\underline{\mu}_{00}^{(1)}
          \underline{A}_0
     \right\}
     (\underline{B}_1)^{-1}.
 \end{eqnarray} 
 \begin{eqnarray}
     \underline{B}_{1} = 
     (\hspace{0.4mm}^t\hspace{-0.4mm}\underline{B}_1)^{-1}
     \left\{
          \underline{\mu}_{00}^{(2)}
        - \underline{A}_0\underline{\mu}_{00}^{(1)}
     \right\}.
 \end{eqnarray} 
 In case the recursion in the block Lanczos algorithm is terminated at
 the $q$th level, the diagonal block element of the Green's function matrix
 can be expanded with the $(2q+1)$th moments, because it is constructed by
 the multiple inverse with the recursion block elements
 $\underline{A}_{n} (n=0\sim q)$, $\underline{B}_{n} (n=1\sim q)$ given by the
 $q$th recursion. As shown in Eqs.~(44) and (45), the recursion block elements
 are expanded in terms of the moments. Thus, $\underline{G}_{00}^L$ contains the
 $0\sim (2q+1)$th moments. This implies that up to $(2q+1)$th moment is
 included in the sum of the moment expansion Eq.~(37), and Eq.~(42) satisfies
 for $r\leq 2q+1$.

 To obtain the moments for the off-diagonal elements of the Green's function
 matrix, multiplying both sides in Eq.~(33) by $(E+0^+)^r$ and integrating
 with respect to the energy $E$, we have
 \begin{eqnarray}
     {\rm Im}
      \int_{-\infty}^{\infty}E^r
       \underline{G}_{0n}^{L}(E+0^+)dE
       = 
     \sum_{m=0}^{n}
    \left(
     {\rm Im}
      \int_{-\infty}^{\infty}E^{r+m}
       \underline{G}_{00}^{L}(E+0^+)dE
    \right)
    \underline{c}_m,
 \end{eqnarray}
 where the block coefficients $\underline{c}_m$ can be written in terms of
 the recursion block elements. As mentioned above the right side of Eq.~(48)
 is equal to the moment of the Hamiltonian for $r+m\leq 2q+1$, so that the
 left side gives the exact moment $\underline{\mu}_{0n}^{(r)}$ for
 $r\leq 2q+1-n$. This means that the off-diagonal elements of the Green's
 function matrix can be expanded with up to the $(2q+1-n)$th moment,
 which results in the expansion of the bond-order $\underline{\Theta}_{0n}^L$ by up to
 the $(2q+1-n)$th moment. Moreover we can relate the bond-orders in the atomic
 basis representation to the moments through the transformation Eq.~(28).
 In the right side of Eq.~(28) the bond-order $\underline{\Theta}_{0q}^L$ for
 $n=q$ determines the maximum order of the moments for the bond-orders based on
 the atomic basis. So we see that the bond-orders in the atomic basis
 representation can be expanded with the moments for $r\leq q+1$.
 Thus, in the block BOP the off-diagonal elements of the Green's function
 matrix can be constructed with the moments for $r\leq q+1$, while the diagonal
 elements have the information of the moments for $r\leq 2q+1$.
 This could be imply the difference in the convergence properties of the bond
 energy and the forces.
 On a simple consideration it is estimated that the rate of the convergence of
 the force is about half as fast as that of the bond energy in terms of
 recursion levels.
 However it should be noted that the contribution of
 $\underline{\Theta}_{0n}^L$ to $\underline{\Theta}_{ij}$ decreases as
 the recursion level $n$ increases, since the Lanczos vectors, which hop
 repeatedly in the atomic connectivity, have their weight away from
 the starting atom as the recursion level $n$ increases. Thus, the
 bond-orders in the atomic basis representation do not have all the
 moments of the higher order more than the $(q+1)$th, but can
 include the higher moments through the $\underline{G}_{0n}^L$ for $n<q$.
 In this case whereas the inexact moments for $r\leq 2q+1-n$ are included in
 the bond-order in the atomic basis representation, the error can be negligible,
 since the bond-orders $\underline{\Theta}_{0n}^L$ become small as 
 the recursion level $n$ increases. So it is stressed that the higher moments can
 be included in the bond-order based on the atomic basis through the Green's
 function $\underline{G}_{0n}^{L}$ for small recursion levels $n$.
 Therefore, it is expected that the forces should be comparable to the bond energy 
 in terms of the convergence rate.
 In Sec.~3 we will discuss this point again numerically.

\subsection{Details on implementation}

 The technical details to implement the block BOP are given in this
 subsection.
 For an infinite system, there could be an infinite number of levels in the
 multiple inverse of the diagonal Green's function. It is often the
 case, however, that the exact values can be replaced by estimated
 values after a certain number of levels, without reducing the
 accuracy significantly. The simplest approximation is to take
 $\underline{A}_n=\underline{A}_{\infty}$,
 $\underline{B}_n=\underline{B}_{\infty}$ for $n>n_{t}$,
 where $n_t$ is the number of exact levels, and $\underline{A}_{\infty}$ and
 $\underline{B}_{\infty}$ are constant block elements. This approximation is
 reasonable from the observation that the scalar elements in both
 $\underline{A}_{n}$ and $\underline{B}_{n}$ converge to constant values or
 oscillate around constant values as n tends to infinity.\cite{Inoue}
 We have only to replace the level for $n=n_t+1$ in the multiple inverse
 with the terminator, since the constant terms can be summed exactly.
 The terminator can be written by a closed form including itself as follows:
 \begin{eqnarray}
   \underline{T}(Z)
        =[Z\underline{{\rm I}}-\underline{A}_{\infty}
            -\hspace{0.4mm}^t\hspace{-0.4mm}\underline{B}_{\infty}
                \underline{T}(Z)
             \underline{B}_{\infty}
         ]^{-1}.
 \end{eqnarray}
 However, this is still a difficult set of equations to solve, so to
 simplify matters we assume that the off-diagonal elements of
 $\underline{T}(Z)$ are zero and all the diagonal elements are the
 same, since the differences between the diagonal elements of
 $\underline{A}_n$ and $\underline{B}_n$ become small
 as the number of the recursion levels increases, respectively.
 Then the identical diagonal element $t(Z)$ of $\underline{T}(Z)$ is written
 as the square root terminator:
 \begin{eqnarray}
    \nonumber
    t(Z) & = & [Z-a-b^2 t(Z)]^{-1}\\
         & = & \frac{1}{b}
              \left[
                 \frac{Z-a}{2b}  
                  -i\sqrt{
                    1- \left(\frac{Z-a}{2b}\right)^2  
                  }
              ~\right],
 \end{eqnarray}
 where $a$ and $b^2$ are given by the means of the diagonal elements of 
 $\underline{A}_{n_t}$ and $\underline{B}^{2}_{n_t}$, respectively.
 Thus, we see that the effect of the terminator is to smear out the sharp
 states with energy $a$ into semielliptical bands. The degree of smearing
 is given by $b$.

 There are two ways to conserve charge neutrality in the
 system: local charge neutrality (LCN)\cite{Sutton} or the total
 charge neutrality (TCN).\cite{Ozaki}
 Within LCN the on-site energies are varied (keeping the splitting
 between on-site $s$ and $p$ energy levels fixed) in order to conserve
 the number of electrons on each atom.
 If the excess charge on site $i$ is $Q_i = Z_i-\sum_{\alpha}N_{i\alpha}$,
 where $Z_i$ is the effective core charge, then the on-site energies can be
 shifted using the response function $X_{i}=\sum_{\alpha}X_{i\alpha}$ for 
 atom $i$ as follows:
 \begin{eqnarray}
    \epsilon_{i\alpha}' = \epsilon_{i\alpha}
                        - \lambda 
                          \frac{Q_i}
                          {X_i},
 \end{eqnarray}
 where $\lambda$ is a parameter to accelerate the convergence, and generally is
 1.0. The response function projected on an atomic orbital $i\alpha$ is given by
 \begin{eqnarray}
    X_{i\alpha} = \frac{2}{\pi} {\rm Im}
           \int 
              [
                G_{i\alpha,i\alpha}(E+i0^+)
              ]^2
              f(
                \frac{E-\mu}{k_BT} 
               )
              dE.
 \end{eqnarray}
 Usually no more than three or four iterations are required to achieve the
 convergence so that the absolute value of $Q/$atom is below $10^{-5}$, since
 $X_{i\alpha}\simeq\partial N_{i\alpha}/\partial \epsilon_{i\alpha}$.
 The assumption of LCN has the advantage that the Madelung energy contribution
 is zero, so that the TB model needs not take this into account in its
 expression for the energy. Also LCN is suitable for parallel computation,
 since the calculations of the bond energy and the forces of each atom are
 perfectly independent within the assumption.
 However, LCN brings an inefficiency in terms of computational effort,
 since LCN requires the Lanczos algorithm to be implemented again,
 after the charge neutralities of all the atoms has been achieved, since
 the recursion block elements are varied by the shift of the on-site energies.
 Thus, the block Lanczos algorithm and the shift of the on-site energies must be
 repeated until self-consistency is accomplished. 
 This self-consistency requires typically twenty iterations.
 This discourages us from applying LCN in the molecular
 dynamics simulations. On the other hand, we can conserve the total number of
 electrons in the system by a shift of the chemical potential in terms of
 TCN. If the excess charge of the system is $Q=\sum_{i}Q_i$, then a good
 approximation of the chemical potential is given by
 \begin{eqnarray}
      \mu' = \mu + \lambda\frac{Q}{X},  
 \end{eqnarray}
 where $X=\sum_i X_i$. The convergence is achieved after only three or four
 iterations. The TCN assumption, corresponding to the micro
 canonical distribution, has physically appropriate meaning, which is
 consistent with the usual electronic structure calculations by 
 diagonalization. Moreover within TCN we need not repeat the Lanczos algorithm,
 since the recursion block elements are not varied by the shift of the chemical
 potential. Thus, TCN has considerable advantage in terms of computational effort.
 The TCN condition reduces the separability of individual atoms
 in the calculations of the band energy and forces, and complicates
 slightly the parallelizability of the program code. However, the evaluation
 and integration of the Green's function, which are time-consuming steps,
 are performed separately. Therefore, we use the TCN constraint to conserve
 the total number of electrons.

 It is required to integrate the Green's functions with the Fermi function
 in order to evaluate the bond energy, bond-orders, and response functions.
 The integration can be carried out in the complex plane by summing up
 an infinite series over the modified Matsubara poles.
 \cite{Horsfield,Horsfield2,Nicholson}
 The general form can be given as follows:
 \begin{eqnarray}
   {\rm Im}\int A(E+i0^+)f(x)dE = 
    -\frac{2\pi}{\beta}{\rm Re}
    \left[
     \mathop{\rm lim}_{P\to \infty}
       \sum_{p=0}^{P-1}
       z_p A(E_p)  
    \right],
 \end{eqnarray}
 with
 \begin{eqnarray}
    E_p = \mu + \frac{2P}{\beta}(z_p-1), \hspace{5mm}
    z_p = {\rm exp}\left(\frac{i\pi(2p+1)}{2P}\right),
 \end{eqnarray}
 where $A(x)$ is an arbitrary function defined in the complex plane, 
 and $\beta=1/k_BT$. Also $E_p$ are the poles of the approximated Fermi
 function in the complex plane. This modified Matsubara summation converges
 rapidly with about 40 complex poles $(P\simeq 40)$ with a high electron
 temperature $(k_BT>0.1~eV)$, although many poles are needed to achieve the
 convergence with a lower electron temperature. In the case of systems with
 a gap between the valence and conduction bands, we need to pay attention to the
 evaluation of the chemical potential, since the response functions
 in the gap become zero as $k_BT$ tends to $0$, so that it is
 difficult to estimate the chemical potential under a low electron
 temperature using Eq.~(53). This can be solved
 by smearing the density of states under a high electron temperature.
 Thus, it is required to evaluate the response functions at 
 high electronic temperatures in order to obtain stable MD
 simulations.

 We now estimate the time-dependence within the block BOP. The total
 system is divided into finite clusters centered on individual atoms
 in order to evaluate the energy and force of each atom. The size of the
 finite cluster is not determined by the size of the total system, but
 by the system and the condition of the MD simulation. Therefore, the
 computational effort is proportional to the number of atoms $N_{\rm atom}$,
 so that the number of computational operations can be written as
 $cN_{\rm atom}$, where $c$ is a proportionality constant. The scaling of
 the constant $c$ can be estimated as a function of the numbers of recursion
 level $q$, atoms within a finite cluster $n_c$, and orbitals
 on an atom $M$. For simplicity it is assumed that the system
 consists of only one type of element with $M$ orbitals.
 In the block Lanczos algorithm the time-consuming step is
 the product of the Hamiltonian matrix by the vector, so that the count
 of operations in the block Lanczos algorithm is nearly proportional to
 $qn_{c}^2M$. 
 At the next step, the inverses and recursive calculations are
 required to evaluate the diagonal and off-diagonal elements of the
 Green's function matrix, respectively, and their integrations are 
 performed as the sum of the residues for the poles in the complex
 plane, so that the count of operations for the evaluations is almost
 proportional to $qPM^3$. Thus, the proportionality constant $c$ can be
 estimated as $c_{L}\times qn_{c}^2M + c_{G}\times qPM^3$, where 
 $c_{L}$ and $c_{G}$ are prefactors of the count of operations
 for the block Lanczos algorithm and the the evaluation of the
 bond-orders, respectively. The prefactors depend on the computer,
 and the system, and the criterion of charge neutrality.
 For example, for the case of a 3 hop cluster, 10 recursion
 levels, and 40 complex poles for diamond carbon, the calculation
 time of the block Lanczos algorithm is comparable to that in
 evaluating and integrating the Green's functions.

 In the remainder of this section the procedure for implementing the
 block BOP is enumerated.
 (I). The partition of the system.
 The hopping range of each atom is determined by terminating the
 system. There are two ways to terminate the system. One of them is
 the physical truncation that the terminated cluster contains atoms
 within a sphere with a certain cutoff radius. 
 The physical truncation can bring inaccurate properties into the
 convergence of the energies, since atoms that have no bonding to
 other atoms can be included in the neighborhood of the cluster
 surface. Moreover, in MD simulations the energies can jump
 discontinuously when an atom moves in or out of the surface of the sphere.
 The more stable way is logical truncation.
 The cluster of size $n$ is here
 defined by all neighbors that can be reached by $n$ hops.
 Provided the cutoff distance for the hopping integral is identical to
 that defining the connectivity of the bonding, the energies are
 continuous as a function of time in MD simulations. Therefore, it is
 desirable to truncate logically the system in terms of accuracy.
 (I\hspace{-0.2mm}I). The block Lanczos algorithm. 
 The Hamiltonians for the individual terminated clusters are
 constructed. For these small cluster Hamiltonians the block Lanczos algorithm
 Eqs.~(14)$\sim$(21) is applied.
 (I\hspace{-0.2mm}I\hspace{-0.2mm}I).
 The evaluations and integrations of the Green's functions.
 In the Lanczos basis representation the diagonal and the off-diagonal
 elements of the Green's functions are evaluated using Eqs~(32) and
 (33), respectively, and then their integrations are performed via the
 modified Matsubara summation with Eq.~(54).
 (I\hspace{-0.2mm}V). The transformation into the atomic basis representation.
 The bond-orders based on the Lanczos basis are transformed into
 those in the atomic basis representation using Eq.~(28).
 (V). The bond energy and forces. 
 From Eqs.~(3) and (9) the bond energy and forces are evaluated,
 respectively.

\section{CONVERGENCE PROPERTIES}

 $O(N)$ methods with linear scaling algorithms are approximate approaches
 compared to the exact diagonalization for dealing with large scale
 systems, so that the realization of the $O(N)$ algorithms is
 accompanied by decreases in computational accuracy in exchange
 for computational efficiency. Therefore, $O(N)$ methods should
 only be applied to atomistic simulations once their accuracy and
 efficiency has been tested.

 In the block BOP two approximations are introduced to reduce
 the computational effort: the number of moments, or recursion
 levels, and the size of the cluster of atoms over which the hops are
 made. The finite approximations for the number of levels and the size
 of the cluster can lead to the errors in the energies and forces.
 Thus, we now investigate the block BOP through several test calculations
 in terms of its accuracy and efficiency.
 In order to ascertain applicable bounds for a wide range of
 materials, the energy and force convergence are examined for an
 insulator (carbon\cite{Xu} in the diamond structure), a semiconductor
 (silicon\cite{Goodwin}), a metal (titanium, described by a canonical
 d-band model), and a molecule (benzene\cite{Horsfield3}) as functions
 of the number of recursion levels and the size of cluster. Moreover,
 in terms of the computational efficiency the block BOP is
 compared with k-space calculations in computational time.
 Also as a test of the quality of the forces, we perform a constant
 energy molecular dynamics (CEMD) simulation of carbon.
 Figure 2 shows the cohesive energy per atom for carbon in the
 diamond structure, silicon in the diamond structure, hcp titanium,
 and benzene. The cohesive energies were calculated using $2\sim 15$
 recursion levels for three, five, and seven shell clusters by the
 logical truncation method, where the three, five, and seven shell clusters
 for the diamond structure include 41, 147, and 363 atoms,
 respectively, and these clusters for the hcp structure contain 153, 587, and
 1483 atoms, respectively. The cohesive energies for carbon and
 silicon converge rapidly to the results of k-space calculations.
 The errors for carbon and silicon are only 1 \% at six recursion
 levels.
 Thus, we see that up to the 13th moment corresponding to six
 recursion levels determine the cohesive energies.
 The contribution of the higher order moments is unimportant,
 since the convergence properties are almost
 identical for three, five, and seven shell clusters. The cohesive
 energy for silicon converges more slowly compared with that of
 carbon in the rate of convergence for the size of cluster.
 This suggests that a semiconductor such as silicon requires
 higher moment than an insulator such as carbon for good convergence
 of the cohesive energy.
 The cohesive energy for the metallic hcp titanium converges
 very quickly in terms of the number of recursion levels.
 For the five and seven shell clusters the cohesive energy converges
 fully to the k-space result, while the convergence value for the
 three shell cluster is in error by 2 \% from the k-space result.
 For benzene the convergence is achieved with a very small cluster (2 shells).
 The error at four recursion levels is only 0.1\%. We see that the
 block BOP can evaluate accurately the cohesive energy for a molecule
 with a sparse structure like benzene, which has both localized
 $\sigma$ bonds and delocalized $\pi$ bonds.

 The calculation of the vacancy formation energy is a severe test
 to distinguish the accuracy of different $O(N)$ methods, since it is a 
 criterion that tests the precision which the dangling bonds caused by the
 vacancy are handled by $O(N)$ method. In practice, the usual
 moment-based $O(N)$ methods fail to reproduce the vacancy formation
 energy of carbon in the diamond structure even when dozens of
 moments are included.\cite{Comparison,Kress}
 The computational error at 30 moments is still about 20 \% compared
 to the k-space result.
 In Fig.~3 we show the vacancy formation energy for carbon in the
 diamond structure, silicon in the diamond structure, and hcp titanium.
 These are calculated as the difference between the energy for a bulk
 unit cell (of 64, 64 or 32 atoms, respectively) with a single atom
 removed, and the perfect bulk cell energy scaled to 63, or 31
 atoms. The results are for an unrelaxed vacancy. The convergence
 properties for carbon and silicon are almost identical. The vacancy
 formation energy in the five and seven shell clusters converges
 smoothly toward the k-space results, while in the 3 shell cluster
 the converged values for carbon and silicon are 15~\%, and 13~\%
 underestimated, respectively. In the seven shell cluster at 15 recursion
 levels the errors for carbon and silicon are only 1\%.
 Thus, we see that the block BOP gives an accurate vacancy
 formation energy for strongly covalent materials such as carbon and
 silicon with the use of about 30 block moments. This remarkable
 result suggests that the block BOP accurately describes 
 dangling bonds in comparison with the usual moment-based methods.
 In Sec.~4 we will discuss the advantages inherent in the block
 BOP by analyzing the vacancy formation energy in terms of 
 different bond-order contributions.
 For titanium the vacancy formation energy converges to the
 k-space result equally within the three, five, and seven shell
 clusters. The error for the 3 shell cluster at 5 recursion levels
 is about 6\%. The vacancy formation energy oscillates with respect to
 the number of recursion levels due to the long range value of the 
 density matrix (see fig.~2 of ref.~23).
 The oscillations are damped by imposing LCN instead of TCN to
 conserve the number of electrons.

 The accuracy of the forces is investigated from two different
 perspectives. The first is the accuracy when compared to the
 exact k-space result, the second is the degree of correspondence
 between the numerical and analytic Hellmann-Feynman forces.
 In order to perform reliable MD simulations the two criteria should
 be satisfied. In Fig.~4 we show the $z$-component of the force on an atom
 in the bulk-terminated (001) surface of carbon, silicon, and hcp
 titanium, and the force on a hydrogen atom on benzene.
 For carbon the force of the three shell cluster overestimates 
 by about 130~\% in comparison with the k-space result, although
 the error in the Hellmann-Feynman term is only 1~\%.
 The forces of the five and seven shell clusters converge 
 smoothly toward the k-space result. The rate of convergence in silicon
 is much better than that of carbon. Even the three shell cluster
 shows a converged value that differs by only 5~\% from the k-space
 result. The three, five, and seven shell clusters of Ti show similar
 convergence properties of the forces, the converged value being
 underestimated by about 8\% compared with the k-space result.
 For benzene the force converges rapidly with small cluster size.
 As discussed in Sec.~2 the bond-orders can be expanded using the
 lower order moments compared with the density of states in the
 block BOP. It can be estimated that the forces should converge more
 slowly at the k-space results than the bond energies, since the
 forces on the atoms are evaluated using the bond-orders. However, these
 numerical results for the forces show that the convergence rate of the
 force is comparable to that of the bond energy. This means that the
 sum of Eq.~(28) converges rapidly as the number of the recursion
 levels increases because of the diffusion of the Lanczos vectors.

 As a test of the consistency between the total energy and the forces,
 constant energy molecular dynamics (CEMD) simulations have been performed
 for carbon. If the forces are equal to the derivative of the total
 energy with respect to atomic positions, the total energy of the
 system is conserved. Thus, the CEMD simulation is a criterion to
 investigate the consistency of forces.
 In Fig.~5 we show the energy for carbon at 1000 and 5000 K as a
 function of time using five and ten recursion levels. The initial
 structure is the diamond lattice, and the unit cell is fixed in volume
 and shape. When the initial temperature of the system is 1000 K,
 the atoms oscillate around the equilibrium positions.
 At five and ten recursion levels we see that the total energy is almost 
 conserved. When the temperature is raised to 5000 K, the carbon
 in the diamond structure transforms into liquid carbon with mainly
 three coordinate structure. From Fig.~5 we see that the forces are of
 good quality at ten recursion levels, while the total energy at
 five recursion levels increases by about 10~eV during the 1 ps,
 which corresponds to a temperature increase of 1800 K. 
 These results indicate that the block BOP can give forces
 consistent with the total energy, provided the proper number of recursion
 levels is used, even for liquid materials such as carbon at a high
 temperature. On the other hand, in the variational DM method, although
 only the Hellmann-Feynman term survives formally as the derivatives of
 the band energy with respect to atomic coordinates, total energy of 
 liquid silicon in the CEMD simulation exhibits a steady upward
 drift.\cite{BowlerDM}
 
 To study the computational efficiency of the block BOP we carry
 out two benchmark tests: the comparison between the block BOP and
 the k-space calculation in computer time, and the relation between
 the computational error and the computer time. Figure~6 shows the
 time to evaluate the energy and forces for a cell containing
 carbon in the diamond structure as function of the number of atoms in
 the cell for the block BOP and k-space using a single
 k-point. The crossover point at which the block BOP becomes
 favorable is about 100 atoms.

 Figures~7(a) and (b) show the relation between the error and
 the the time per atom to evaluate the energy and forces in the
 calculations of the vacancy formation energy of diamond carbon
 and hcp titanium, respectively. We see that the block BOP can
 calculate the vacancy formation energy to high accuracy within
 almost the same computational time as the other moment-based results
 reported by D.R.Bowler et al.\cite{Comparison} where the calculations
 were performed using the same computational facilities.
 We note that the block BOP has given a good convergent
 result of the vacancy formation energy in diamond carbon for the
 first time with a moments-based method, while the computational
 time to achieve this convergence is still ten times slower than that
 of the DM method. This work, therefore, still supports the conclusions
 of the study in ref.~23 that the DMM is best for
 systems with energy gaps, but that moments-based methods such as
 BOP are best for metallic systems.

\section{ANALYSIS OF VACANCY FORMATION ENERGY}

 The block BOP can provide chemical insight into the nature of the
 bonding in molecules and solids in terms of the bond-order.
 The bond-order is an useful quantity indicating the strength of
 bonding between two atoms.
 In practice, it is well known that the bond length is nearly proportional
 to the bond-order for the $\pi$ bonded hydrocarbons.\cite{Warshel}

 In this section we analyze the vacancy formation energy
 of carbon in the diamond structure in terms of the bond-order, and
 discuss the reason why the usual moment-based methods can not
 reproduce the vacancy formation energy even with several dozens of moments.
 \cite{Comparison,Kress}  The reduced TB
 model\cite{RTB,ABOP,ABOP2,Conrad} is introduced in
 order to clarify the analysis of the vacancy formation
 energy in terms of the two scalar bond-orders $\Theta_{\sigma}$ and 
 $\Theta_{\pi}$, respectively.
 In the reduced TB model the three independent bond integrals $h_{ss\sigma}$,
 $h_{pp\sigma}$, and $h_{sp\sigma}$ are reduced to the two independent
 variables $h_{\sigma}$ and $p_{\sigma}$ by assuming that
 $h_{sp\sigma}$ is the geometric mean of $\vert h_{ss\sigma}\vert$ and
 $h_{pp\sigma}$. This allows the $\sigma$ bond energy between atoms
 $i$, and $j$ to be written as the single quantities
 $h_{\sigma}(R_{ij})\Theta_{\sigma}^{ij}$ rather than the sum of
 three terms
 $h_{ss\sigma}\Theta_{ss\sigma}$,
 $2h_{sp\sigma}\Theta_{sp\sigma}$, and 
 $h_{pp\sigma}\Theta_{pp\sigma}$, respectively.
 That is, we can write
 \begin{eqnarray}
    \nonumber
    E_{\rm bond}^{(ij)} & = & E_{\sigma}^{(ij)} + E_{\pi}^{(ij)}\\
                    & = &  -2 h_{\sigma}^{(ij)}\Theta_{\sigma}^{(ij)}
                           -4 h_{\pi}^{(ij)}\Theta_{\pi}^{(ij)},   
 \end{eqnarray}
 where 
 \begin{eqnarray}
    \nonumber
    h_{\sigma} & = & (1+p_{\sigma})\vert h_{ss\sigma}\vert,\\[1mm]
    \Theta_{\sigma} & = &
         \frac{\theta_{ss\sigma}+2\sqrt{p_{\sigma}}
               \theta_{sp\sigma}+p_{\sigma}\theta_{pp\sigma}}
               {1+p_{\sigma}},
 \end{eqnarray}
 with $p_{\sigma}=h_{pp\sigma}/\vert h_{ss\sigma}\vert$. 
 All the hopping integrals and bond-orders are defined
 as positive quantities. 
 In addition, the cutoff distance in the hopping integrals and
 repulsive potential is reduced from 2.6~\AA~in the original
 TB fit\cite{Xu} to 2.5~\AA. This modification simplifies the analysis,
 because atoms on the diamond lattice do not interact with this second
 neighbors who lie at a distance of 2.517~\AA.
 Also we apply LCN to the analysis rather than TCN, since chemical concepts
 like the promotion energy require the total number of electrons on a given
 atom to be invariant as the atoms we brought together to form the
 bond. In table I we give the cohesive energy and vacancy formation
 energy of carbon in the diamond structure calculated using the
 original TB and the reduced TB methods. The changes in the cohesive and
 vacancy formation energies introduced by the reduced TB simplifications
 are only 0.1~\% and 3~\%, respectively. Therefore, it is an excellent
 approximation to analyze the vacancy formation energy using the reduced
 TB method with LCN.

 Figure~8 shows the diamond lattice with a vacancy. There are four
 first (FN) and twelve second (SN) neighboring atoms about the vacancy.
 The twenty four third neighboring atoms in total are grouped into 
 two kinds of atoms (TN, TN$'$), each of them including twelve atoms.
 The number of valence $s$ and $p$ electrons and the corresponding
 promotion energy of the FN, SN, TN, and TN$'$ atoms are given in table II.
 The number of valence $s$ electrons on the FN atom increases by
 about 6~\% compared with that of a carbon atom in the perfect
 structure, which corresponds to an increase of 0.27 electrons in total
 over the four FN atoms. This increase in the number of $s$ electrons
 on the FN sites reflects that the $s$ component of the dangling bond
 is attracted firmly at the core of the carbon.
 The number of $s$ electrons on the SN, TN, TN$'$ atoms, on the other
 hand, is almost the same as that of the perfect structure.
 We see, therefore, from table II that 97 \% of the total change
 in promotion energy resides in the FN shell of atoms about 
 the vacancy, so that the redistribution of $s$ and $p$ valence
 electrons occurs mainly within the first shell.
 The change in the promotion energy stabilizes the vacancy by 
 1.858~eV.

 Table III shows the bond-orders and bond energies for 
 $\sigma$ and $\pi$ bonds between the pairs of atoms FN-SN, SN-TN, and
 SN-TN$'$, respectively, in the presence of a vacancy.
 The $\sigma$ bond-order for FN-SN increases by 0.4~\%, whereas for SN-TN
 and SN-TN$'$ bonds it decreases by 0.6~\% and 0.1~\%
 respectively compared with that of the perfect structure.
 This oscillatory behavior in the variation of the bond-orders
 reflects the screening of the vacancy. However, the very small variation
 in the $\sigma$ bond-order reflects the localized nature of the 
 $\sigma$ bonding in carbon which is a strongly covalent material. 
 For the $\pi$ bonding the bond-order between FN and SN atoms
 increases by 36~\% compared with that of the perfect structure.
 This increase means that the $p$ electron of the dangling bond
 participates in the $\pi$ bonding between FN and SN atoms rather than
 being attracted solely to the core of the carbon vacancy.
 If we had assumed that the bond-orders are invarint to the formation of 
 a vacancy, then the vacancy formation energy would have been overestimated
 by 1.752~eV. This additional stabilization energy to the formation of 
 the vacancy is distributed between the $\sigma$ and $\pi$
 bond energies, as -0.799 and 2.551~eV contributions, respectively.
 Thus, the absolute ratio of the $\sigma$ to $\pi$ contributions
 is about 1 to 3, which is considerably larger than the ratio of
 the $\sigma$ and $\pi$ bonding energy (18.054:0.635) in the perfect
 diamond lattice. In the $\sigma$ bond energy the contribution
 of the SN-TN and SN-TN$'$ bonding to this stabilization energy is
 comparable to that of the FN-SN bonding.
 On the other hand, the stabilization energy for the $\pi$ bonding
 comes from mainly the FN-SN bonding as the $p$ electron in the
 dangling bond only participates in the $\pi$ bonding between the
 FN and SN atoms. 
 Finally, the total vacancy formation energy (9.783~eV) can be separated
 as the difference of the repulsive energy (-23.983~eV), the bond 
 energy of the absent bonds reproduced by the vacancy (37.380~eV),
 the stabilization of the promotion energy (-1.858~eV),
 and the stabilization of the bond energy (-1.752~eV).

 Figure~9 shows the errors in the bond-orders for
 $\sigma$ and $\pi$ bonds between the FN and SN atoms.
 We see that the rate of convergence with respect to the number of
 recursion levels of the $\pi$ bond-order is twice as large as
 that of the $\sigma$ bond-order.

 Thus, we have found that the block BOP can separate the different
 behavior of $\sigma$ and $\pi$ orbitals correctly. In particular, it can 
 reproduce the different magnitude of reconstruction for the
 vacancy and the convergence rate with respect to the number of the
 recursion levels. In the usual scalar moment-based methods the
 $\sigma$ and $\pi$ orbitals are not separated, since an averaged
 moment is used for the two kinds of orbitals.\cite{Aoki}
 This means that the different properties of the $s$ and $p$
 electrons in the dangling bond are averaged with respect to 
 the vacancy formation energy and the convergence rate.
 As a result a great many moments are required in order to reproduce
 the vacancy formation energy in the usual scalar moment-based method.

\section{LARGE SCALE SIMULATION}

 The block BOP is applicable to the atomistic simulations
 of large systems including thousands of atoms.
 In this section we discuss the parallel computation required to perform
 such large scale atomistic simulations and illustrate the method
 with an application to the deformation of a single-wall carbon nanotube
 under compression. It is very easy to parallelize the program code because
 of the highly independent nature of the algorithm which evaluates
 the energy and force for each atom. We have only to parallelize essentially
 the three main loops in the program code: the block Lanczos transformation, 
 the determination of the number of electrons on each atom, 
 and the evaluation of forces. In these loops independent
 calculations can be performed for each atom, since no information needs
 to be passed between the individual atoms.
 The majority of the computational effort is occupied
 by the calculations in these three loops. Thus, if the three loops are
 parallelized, we have almost the ideal parallel code.
 Figure~10 shows the time to evaluate the energies and forces of 
 a diamond unit cell including 512 atoms as a function of the number
 of processors. It is found that the scaleability is almost ideal.
 The parallelization was done using an automatic parallel compiler.
 The ideal scaleability brought by the use of the automatic parallel
 compiler indicates the simplicity of the algorithms within block BOP.

 We have performed a large scale atomistic simulation for the deformation
 of a single-walled (10,10) nanotube under compression along the shaft
 as an application of the block BOP.
 The nanotube, which has a length of 140~\AA , includes 2280
 carbon atoms. The compression along the shaft was
 performed by the following geometric optimization with a constraint.
 First, the $z$-coordinate of all the atoms in the nanotube oriented along
 the $z$-axis are scaled as the length of shaft decreases 0.1~\% of its
 initial length. Second, the scaled structure is optimized
 geometrically with a constraint that the $z$-coordinate of the atoms
 within 7~\AA~ of both ends are kept fixed.
 By applying repeatedly the optimization to the nanotube, the shaft of
 the nanotube can be compressed statically. 
 In the early stage of the compression the nanotube shrinks,
 maintaining the shape. However, the nanotube buckles periodically
 when the length of shaft reaches about 80~\% of the initial
 structure. Figure~11 shows a snapshot of the ripple-buckling
 nanotube. The mean wavelength of the ripple-buckling is 4.8~\AA.
 The appearance of the ripple-buckling is very similar to the behavior
 observed by transmission electron microscope (TEM) and atomic force
 microscope (AFM) measurements.\cite{Iijima,Falvo}
 A detail of discussion of the deformation and elastic
 properties of carbon nanotubes will be presented elsewhere.

\section{CONCLUSIONS}

 We have presented the theory of the block BOP based on the
 Lanczos basis representation and the block Lanczos algorithm with a
 single site as the starting state within the orthogonal tight-binding
 representation. The efficient $O(N)$ algorithm provides a general recursion
 method for evaluating the bond energy and forces. In the Lanczos
 basis representation the off-diagonal block elements of Green's
 function matrix can be related to the diagonal block elements through
 a simple recurrence relation. As a result the bond-orders can be
 easily evaluated.
 From the convergence properties for the bond energies
 and forces it is found that the method is applicable to a wide rage
 of materials (insulators, semiconductors, metals, and molecules) with
 a considerable reduction in the computational effort compared to 
 k-space methods.
 The algorithm becomes more efficient than the k-space calculation
 when the number of atoms exceeds about 100. Constant energy
 molecular dynamics simulations for carbon show that the forces are
 consistent with the total energy, even if the method is
 applied to liquids.
 Moreover, the use of the block Lanczos algorithm guarantees that
 block BOP represents the different properties of the $\sigma$ and
 $\pi$ bonds correctly, so that the vacancy formation energy of 
 diamond is reproduced correctly for the first time by a moments-based
 method. Finally, block BOP is very easy to parallelize,
 the parallelization of the three main loops giving almost the ideal
 scaleability. The deformation of carbon nanotube under compression
 was demonstrated as an application of the method to large scale
 atomistic simulations. Thus, we conclude that the block BOP is an
 efficient $O(N)$ method to perform large scale atomistic
 simulations of a wide variety of materials.

\acknowledgments

  We would like to thank A.~P.~Horsfield, I.~I.~Oleinik, D.~Nguyen-Manh,
  and H.~Nakamura for many useful discussions and many comments.
  T.~O. gratefully acknowledges many encouragements and enlightening
  discussions with T.~Mitani and Y.~Iwasa.
  T.~O. would like to thank the Department of Materials at 
  Oxford University for their hospitality during the fall of 1998.
  M.~A. and D.~G.~P. would like to thank the British Council for 
  their support.
  Part of the computation in this work has been done using
  the computational facilities of the Material Modelling Laboratory
  (MML) in the Department of Materials at Oxford University.

\appendix

 \section{}

 We derive a novel method for evaluating the exact force,
 which is consistent with the total energy, at any level of
 approximation.
 In this method the derivatives of the diagonal Green's functions can
 be evaluated indirectly by making use of the Lanczos basis
 representation, while the method is similar to the global density of
 states (GDOS) method\cite{GDOS} as regards the diagonal Green's
 functions are differentiated. The contribution from the band energy
 to the force is written using Eqs.~(2) and (5) as follows:
 \begin{eqnarray}
   \nonumber
   {\bf F}_k^{({\rm band})} & = & 
                    -\frac{\partial E_{\rm band}}{\partial {\bf r}_k}\\
                      & = &
                     \frac{2}{\pi}{\rm Im}\int
               {\rm tr}\left\{
                          \frac{\partial G(E+{\rm i0^+})}{\partial {\bf r}_k}
                       \right\}E f(x) dE,
  \end{eqnarray}
 The trace of the right side in Eq.~(A1) can be divided into the
 diagonal block elements for individual atoms:
 \begin{eqnarray}
   \nonumber
  {\rm tr}\left\{
               \frac{\partial G(E+{\rm i0^+})}{\partial {\bf r}_k}
          \right\} 
           & = &
       \sum_i
  {\rm tr}\left\{
         \frac{\partial \underline{G}_{ii}(E+{\rm i0^+})}
              {\partial {\bf r}_k}
         \right\}\\
           & = &
       \sum_i
  {\rm tr}\left\{
         \frac{\partial \underline{G}_{00}^{L^{(i)}}(E+{\rm i0^+})}
              {\partial {\bf r}_k}
         \right\},
 \end{eqnarray}
 where the index $L^{(i)}$ represents the representation based on 
 the Lanczos basis with atom $i$ as the starting site. In the Laczos basis
 representation the off-diagonal block elements of the Green's
 function matrix are evaluated through the recurrence relation
 Eq.~(33) which is derived from the identity
 $(Z{\rm I}-H)G(Z)={\rm I}$.
 Thus, the following useful relation can be derived for the derivative
 of the Green's function:
 \begin{eqnarray}
    \frac{\partial G^{L^{(i)}}(Z)}{\partial {\bf r}_k} =
            G^{L^{(i)}}(Z)
    \frac{\partial H^{L^{(i)}}}{\partial {\bf r}_k}
            G^{L^{(i)}}(Z),
 \end{eqnarray}
 Taking account of the (0,0) block element of both sides in Eq.~(A3)
 we have
 \begin{eqnarray}
     {\rm tr}\left\{
        \frac{\partial \underline{G}_{00}^{L^{(i)}}(Z)}
             {\partial {\bf r}_k}
        \right\}
        = 
          \sum_{mn}
          {\rm tr}
          \left\{
            \underline{G}_{n0}^{L^{(i)}}(Z)  
            \underline{G}_{0m}^{L^{(i)}}(Z)
            \frac{\partial \underline{H}_{mn}^{L^{(i)}}}
                 {\partial {\bf r}_k}
         \right\},
 \end{eqnarray}
 Substituting Eqs.~(A2) and (A4) for the trace in Eq.~(A1)
 we can derive exactly the contribution from the band energy to the
 force at any level of approximation as the following very
 compact form:
 \begin{eqnarray}
   \nonumber
   {\bf F}_k^{({\rm band})} & = & 
         -\sum_{i}\sum_{mn}
           {\rm tr}\left\{
            \underline{\tau}_{nm}^{L^{(i)}}       
            \frac{\partial \underline{H}_{mn}^{L^{(i)}}}
                 {\partial {\bf r}_k}
           \right\}\\
                      & = &
         -\sum_{i}\sum_{m\alpha,n\beta}
            \tau_{n\beta,m\alpha}^{L^{(i)}}           
            \frac{\partial H_{m\alpha,n\beta}^{L^{(i)}}}
                 {\partial {\bf r}_k},
  \end{eqnarray}
  where
 \begin{eqnarray}
      \underline{\tau}_{nm}^{L^{(i)}} =
        -\frac{2}{\pi}{\rm Im}\int
           E
           \underline{G}_{n0}^{L^{(i)}}(E+{\rm i}0^+)
           \underline{G}_{0m}^{L^{(i)}}(E+{\rm i}0^+)
           f(x)dE,
 \end{eqnarray}
 where the summation for atom $i$ is taken for all the atoms which
 include the atom $k$ in the truncation of system.
 This expression for forces has a very similar form to the
 Hellmann-Feynman force (see Eq.~(9)). However $\tau$ is a new
 dimensionless quantity indicating the strength of bonding between
 two Lanczos vectors.

 A simple example is shown. Consider the force on atom $1$ being 
 at the end of a linear s-valent trimer along the $x$-axis. It is assumed 
 that the trimer has two electrons and the hopping integral between
 the pairs of atoms 1-2, 2-3, and 1-3 are $-h_1$, $-h_2$,
 and zero, respectively, where the $h_1$ and $h_2$ are positive and
 $h_2<h_1$, and also on-site energies for all the atoms are zero.
 If the recursion is approximated at the first levels, then the
 Green's functions are given as follows:
 \begin{eqnarray}
    G_{00}^{L^{(1)}}(Z)&=& \frac{\frac{1}{2}}{Z-h_1}
                         +\frac{\frac{1}{2}}{Z+h_1}.\\
    G_{01}^{L^{(1)}}(Z)&=& \frac{\frac{1}{2}}{Z-h_1}
                         +\frac{\frac{-1}{2}}{Z+h_1}.\\
    G_{00}^{L^{(2)}}(Z)&=& \frac{\frac{1}{2}}{Z-\sqrt{h_1^2+h_2^2}}
                         +\frac{\frac{1}{2}}{Z+\sqrt{h_1^2+h_2^2}}.\\
    G_{01}^{L^{(2)}}(Z)&=& \frac{\frac{1}{2}}{Z-\sqrt{h_1^2+h_2^2}}
                         +\frac{\frac{-1}{2}}{Z+\sqrt{h_1^2+h_2^2}}.\\
    G_{00}^{L^{(3)}}(Z)&=& \frac{\frac{1}{2}}{Z-h_2}
                         +\frac{\frac{1}{2}}{Z+h_2}.\\
    G_{01}^{L^{(3)}}(Z)&=& \frac{\frac{1}{2}}{Z-h_2}
                         +\frac{\frac{-1}{2}}{Z+h_2}.
 \end{eqnarray}
 The band energy of the trimer can be evaluated from the residues of
 the poles, which are occupied, in the diagonal elements of the Green's
 functions, namely 
 \begin{eqnarray}
    E_{\rm band} = -h_1 - \sqrt{h_1^2+h_2^2}.
 \end{eqnarray}
 In the force evaluation by Eq.~(A5) we have only to calculate
 the $\tau_{01}^{L^{(1)}}$(or $\tau_{01}^{L^{(1)}}$), and
 $\tau_{01}^{L^{(2)}}$(or $\tau_{10}^{L^{(2)}}$), since the
 derivatives of Hamiltonian with respect to the position of
 atom $1$ corresponding to other $\tau$'s are zero.
 The integrands for these $\tau$'s are 
 \begin{eqnarray}
   \nonumber
   \lefteqn{ZG_{00}^{L^{(1)}}(Z)G_{01}^{L^{(1)}}(Z)}\\
   \nonumber
                  && = \frac{\frac{1}{4}}{Z-h_1}
                      +\frac{\frac{-1}{4}}{Z+h_1}
                      +\frac{\frac{1}{4}h_1}{(Z-h_1)^2} 
                      +\frac{\frac{1}{4}h_1}{(Z+h_1)^2},\\\\
   \nonumber
   \lefteqn{ZG_{00}^{L^{(2)}}(Z)G_{01}^{L^{(2)}}(Z)}\\
   \nonumber
       && = \frac{\frac{1}{4}}{Z-\sqrt{h_1^2+h_2^2}}
           +\frac{\frac{-1}{4}}{Z+\sqrt{h_1^2+h_2^2}}\\
       &&\hspace{5mm}
  +\frac{\frac{1}{4}\sqrt{h_1^2+h_2^2}}{\left(Z-\sqrt{h_1^2+h_2^2}\right)^2} 
  +\frac{\frac{1}{4}\sqrt{h_1^2+h_2^2}}{\left(Z+\sqrt{h_1^2+h_2^2}\right)^2}.
 \end{eqnarray}
 From the residues of the poles which are occupied
 we see that $\tau_{01}^{L^{(1)}}(=\tau_{10}^{L^{(1)}})$,
 and $\tau_{01}^{L^{(2)}}(=\tau_{01}^{L^{(2)}})$ are $-1/2$ and
 $-1/2$, respectively.
 Transforming the derivative of the Hamiltonian by Eq.~(27) into the 
 the Laczos basis representation we have 
 \begin{eqnarray}
   \frac{\partial H_{01}^{L^{(1)}}}{\partial x_1} &=& 
   \frac{\partial H_{10}^{L^{(1)}}}{\partial x_1}  =
   \frac{\partial h_1}{\partial x_1},\\
   \frac{\partial H_{01}^{L^{(2)}}}{\partial x_1} &=& 
   \frac{\partial H_{10}^{L^{(2)}}}{\partial x_1}  =
   \frac{h_1}{\sqrt{h_1^2+h_2^2}}
   \frac{\partial h_1}{\partial x_1}.
 \end{eqnarray}
 Thus, we can evaluate the contribution from the band energy to the
 force on atom $1$ using Eq.~(A5), namely
 \begin{eqnarray}
  F_{x_1}^{({\rm band})}
   \nonumber
   &=&
  -2\tau_{01}^{L^{(1)}}\frac{\partial H_{10}^{L^{(1)}}}{\partial x_1}
  -2\tau_{01}^{L^{(2)}}\frac{\partial H_{10}^{L^{(2)}}}{\partial x_1}\\
   &=&
  \nonumber 
  -2\times\frac{-1}{2}\times \frac{\partial h_1}{\partial x_1}
  -2\times\frac{-1}{2}\times \frac{h_1}{\sqrt{h_1^2+h_2^2}}
                             \frac{\partial h_1}{\partial x_1}\\
   &=&
   \frac{\partial h_1}{\partial x_1}+\frac{h_1}{\sqrt{h_1^2+h_2^2}}
                                     \frac{\partial h_1}{\partial x_1}.  
 \end{eqnarray}
 The result is identical to the contribution from the derivative of
 Eq.~(A13) with respect to the $x$-coordinate of atom $1$. We see that
 the force by Eq.~(A5) is certainly exact.

%

 \begin{figure}
   \caption{
          The Lanczos vectors on the s-valent square lattice.
          a), b), and c) are an initial state $\vert L_0\rangle$, 
          $\vert L_1\rangle$, and $\vert L_2\rangle$, respectively.
          The diameter of the circles is proportional to the magnitude of
          the expansion coefficient in the Lanczos vector.}
 \end{figure}

 \begin{figure}
  \caption{
    The cohesive energy for carbon in the diamond structure,
    silicon in the diamond structure, hcp titanium, and benzene as 
    a function of number of recursion levels for three, five, and 
    seven shell clusters, calculated using a square root terminator,
    and $k_BT=0.1~$eV.}
 \end{figure}

 \begin{figure}
  \caption{
    The vacancy formation energy for carbon in the diamond structure,
    silicon in the diamond structure, and hcp titanium for three, five,
    and seven shell clusters as a function of number of recursion
    levels, calculated using a square root terminator, a total
    charge neutrality, and $k_BT=0.1~$eV.}
 \end{figure}

 \begin{figure}
  \caption{
    The $z$-component of the force on an atom on the carbon (001)
    surface, silicon (001) surface, titanium (001) surface, and on
    a hydrogen atom in benzene for three, five, and seven shell
    clusters as a function of number of recursion levels, calculated
    using a square root terminator, total charge neutrality,
    and $k_BT=0.1~$eV.}
 \end{figure}

 \begin{figure}
  \caption{
   The potential, kinetic, and total energies as a function of time for
   molecular dynamics simulations of carbon using a three hop
   logically truncated cluster, a square root terminator, total charge
   neutrality, and $k_BT=0.1$~eV. In panels (a) and (b) the results
   are for five and ten recursion levels at 1000 K, respectively,
   whereas in panels (c) and (d) they are for five and ten recursion
   levels at 5000 K, respectively. The time step is 0.5 fs.}
 \end{figure}

 \begin{figure}
  \caption{
    The time to perform the energy and the force evaluation for carbon
    in the diamond structure as a function of number of atoms in the
    cell for the block BOP, calculated using a three hop logically
    truncated cluster, and k-space. The calculations were performed on
    an IBM RS/6000 workstation.}
 \end{figure}

 \begin{figure}
  \caption{
    The error in the carbon (a) and titanium (b) vacancy formation
    energies against the time taken per MD step per atom for three,
    five, and seven shell clusters. The calculations were carried
    out with a square root terminator, total charge neutrality,
    and $k_BT=0.1$~eV on a HP9000/735 workstation.}
 \end{figure}

 \begin{figure}
  \caption{
     Diamond lattice with a vacancy. FN and SN label the first and
     second neighboring atoms to the vacancy, respectively.
     Two kinds of third neighboring atoms are
     distinguished by TN and TN$'$.}
 \end{figure}

 \begin{figure}
  \caption{
    The errors in the bond-order for $\sigma$ and $\pi$ bonds between
    the FN and SN atoms, where the 15 recursion level bond-orders
    $\Theta_{\sigma}=0.9156$ and $\Theta_{\pi}=0.1412$ were taken as
    the exact values.}
 \end{figure}

 \begin{figure}
  \caption{
           The calculation time to evaluate the energy and the forces
           of a 512 atom carbon cell as a function of the number of
           processors by the parallel code. The benchmarks were
           performed on a Sun Ultra 10000 StarFire, using a three shell
           cluster, five recursion levels, and a square root
           terminator.}
 \end{figure}

 \begin{figure}
  \caption{
       Ripple-buckling single-wall (10,10) nanotube under
       compression along the shaft. The snapshot is at 80~\% of
       the initial length (140~\AA ). The calculation was performed
       with a five hop logically truncated cluster, ten recursion
       levels, a square root terminator, total charge neutrality,
       and $k_BT=0.1$~eV. Within the calculation conditions 
       the error is about 1~\% in the bond energy for the initial
       structure compared with the k-space result.}
 \end{figure}

 \begin{table}
 \caption{
    Comparison of the original and the reduced TB method with respect
    to the predicted cohesive energies of carbon in the perfect diamond
    structure and the diamond unit cell including 64 atoms with a single
    atom removed.
    The calculations were performed with a logical truncation of
    a 7 shell cluster, 15 recursion levels, a square root terminator, 
    local charge neutrality, and $k_BT=0.1$~eV.}
  \begin{tabular}{lccc}
     \hline
                   &   Perfect  &  Vacancy   &{\scriptsize Vacancy Formation}\\
                   &  (eV/atom) & (eV/atom)  &{\scriptsize Energy (eV)}       \\
     \hline
  {\bf Original} 
                   &            &            &                  \\
   K-Space         &   -7.251   &  -7.091    &    10.110        \\
     BOP           &   -7.249   &  -7.090    &    10.004        \\
    \hline
  {\bf Reduced}
                   &            &            &                  \\
   K-Space         &   -7.254   &  -7.098    &     9.860        \\
     BOP           &   -7.256   &  -7.100    &     9.783        \\
     \hline
  \end{tabular}
 \end{table}

\newpage

 \begin{table}
 \caption{
          The number of valence $s$ and $p$ electrons and the promotion
          energies of the first (FN), second (SN), and third (TN, TN$'$) neighboring 
          carbon atoms about a vacancy in the diamond structure.
          $\Delta E_{\rm prom}$ is defined as the difference
          between the structure with a vacancy and the perfect structure 
          in the promotion energy.}
      \begin{tabular}{cccccccc}
         \hline
                                  &  Total    & Perfect&   FN   &   SN  &  TN    &  TN$'$ & Others\\
         \hline
         $        N_s      $      &   -       & 1.203  & 1.271  & 1.203 & 1.202 & 1.205 &- \\
         $        N_p      $      &   -       & 2.797  & 2.729  & 2.797 & 2.798 & 2.795 &- \\
         $E_{\rm prom}({\rm eV}/atom)$ &  -   & 5.334  & 4.886  & 5.342 & 5.344 & 5.328 &-\\
      $\Delta E_{\rm prom}({\rm eV}/atom)$ & -&  -     &-0.448  & 0.008 & 0.010 &-0.006 &-\\
         \hline
          Number of atoms         &    63     & -       &    4  &  12   &  12   &  12   & 23\\
      $\Delta E_{\rm prom}({\rm eV})$ & -1.858& -       &-1.793 & 0.097 & 0.115 &-0.076 &-0.201 \\
         $\frac{\Delta E_{\rm prom}}{\hspace{0.5mm}\Delta E_{\rm prom}({\rm Total})}
         \times 100(\%)$          &  100      & -       & 96.5  & -5.2  &-6.2   &  4.1  & 10.8  \\
         \hline
      \end{tabular}
 \end{table}

\newpage

 \begin{table}
 \caption{
      The bond-orders and the bond energies for $\sigma$ and $\pi$
      bonds between the pairs of atoms FN-SN, SN-TN, and
      SN-TN$'$, respectively, in the presense of a vacancy.
      In the calculations of the bond energies
      $E_{\sigma}$ and $E_{\pi}$, the hopping integrals are
      $h_{\sigma}=9.903$ eV and $h_{\pi}=1.533 $ eV.
      $\Delta E_{\sigma}$ and $\Delta E_{\pi}$ are defined as 
      the difference with the bond energy between the structure with
      a vacancy and the perfect structure.
      $\Delta E_{\sigma+\pi}$ represents
      $\Delta E_{\sigma}+\Delta E_{\pi}$.}
    \begin{tabular}{ccccccc}
       \hline
                             & Total    & Ideal   &  FN-SN   &  SN-TN & SN-TN$'$   & Others \\
       \hline
      $\Theta_{\sigma}$      &   -      & 0.9116  & 0.9161   & 0.9058 & 0.9107   &  -   \\
      $\Theta_{\pi}$         &   -      & 0.1036  & 0.1412   & 0.1034 & 0.1020   &  -   \\
      $E_{\sigma}$(eV/bond)  &   -      &-18.055  &-18.143   &-17.941 &-18.036   &  -   \\
      $E_{\pi}$(eV/bond)     &   -      &-0.635   &-0.866    &-0.634  &-0.625    &  -  \\
      \hline
        Number of bonds      &   124    &   124   &   12     &   12   &  24     &  76 \\
    $E_{\sigma}$(eV)         &-2237.918 &-2238.717&-217.718  &-215.288&-432.866 &-1372.046\\
    $E_{\pi}$(eV)            &-81.291   &-78.740  &-10.386   &-7.605  &-15.000  &-48.300  \\
    $E_{\sigma+\pi}$(eV)     &-2319.209&-2317.457 &-228.105  &-222.893&-447.866 &-1420.362\\
      $\Delta E_{\sigma}$    &  0.799  &   -      &-1.068    &  1.362 &0.434    &  0.071 \\
      $\Delta E_{\pi}$       &-2.551   &   -      &-2.766    &  0.015 &0.240    & -0.040 \\
      $\Delta E_{\sigma+\pi}$&-1.752   &   -      &-3.835    &  1.377 &0.674    & -0.032 \\
      $\frac{\Delta E_{\sigma}}{\Delta E_{\sigma+\pi}({\rm Total})}
       \times 100(\%)$       &-45.6    &   -      & 61.0     &-77.7   &-24.8    &-4.1      \\
      $\frac{\Delta E_{\pi}}{\Delta E_{\sigma+\pi}({\rm Total})}
       \times 100(\%)$       &145.6    &   -      &157.9     &-0.9    &-13.7    & 2.3     \\
      $\frac{\Delta E_{\sigma+\pi}}{\Delta E_{\sigma+\pi}({\rm Total})}
       \times 100(\%)$       &  100    &   -      &218.9     &-78.6   & -38.5   & 1.8 \\
       \hline
    \end{tabular}
 \end{table}

\end{document}